\documentclass[epj]{svjour}
\usepackage{graphics}

\begin{document}
\title{Enhancement of stability in randomly switching potential with metastable state}

\author{B. Spagnolo\inst{1}, A. A. Dubkov\inst{2}, \and N. V. Agudov\inst{2}}

\institute{INFM, Unit\`{a} di Palermo and Dipartimento di Fisica e
Tecnologie Relative - Universit\`{a} di Palermo, Viale delle
Science, pad.18, I-90128 Palermo, Italia \email{spagnolo@unipa.it}
\and Radiophysics Department, Nizhny Novgorod State University -
23 Gagarin Ave., Nizhny Novgorod, 603950 Russia
\email{dubkov@rf.unn.ru, agudov@rf.unn.ru}} \offprints{Bernardo
Spagnolo}

\date{Received: date / Revised version: date}

\abstract{ The overdamped motion of a Brownian particle in
randomly switching piece-wise metastable linear potential shows
noise enhanced stability (NES): the noise stabilizes the
metastable system and the system remains in this state for a
longer time than in the absence of white noise. The mean first
passage time (MFPT) has a maximum at a finite value of white noise
intensity. The analytical expression of MFPT in terms of the white
noise intensity, the parameters of the potential barrier, and of
the dichotomous noise is derived. The conditions for the NES
phenomenon and the parameter region where the effect can be
observed are obtained. The mean first passage time behaviours as a
function of the mean flipping rate of the potential for unstable
and metastable initial configurations are also analyzed. We
observe the resonant activation phenomenon for initial metastable
configuration of the potential profile.
}
\authorrunning{B. Spagnolo, A. A. Dubkov, N. V. Agudov}
\titlerunning{Enhancement of stability in randomly switching metastable state}
\maketitle
\section{Introduction}
\label{intro}

The thermally activated escape from a metastable state in
fluctuating potential is of great importance to many natural
systems, ranging from physical and chemical systems to biological
complex systems. A particular challenging direction are systems
far from thermal equilibrium, either due to non-thermal noise or
external deterministic periodical forces
\cite{Rei02,Sme99,Leh00,Mai01,Hir82,Day92,Man96,Man98,Man00,Agu01,Agu00}.
A typical problem is the enhancement of stability of metastable
and unstable states due to the external noise
\cite{Man96,Man98,Man00,Agu01,Agu00}. The noise-enhanced stability
(NES) phenomenon was observed experimentally and numerically in
various physical systems (see Refs.
\cite{Hir82,Day92,Man96,Man98,Man00,Agu01,Agu00,Agu03,Fia03,Mal96}
and, as a recent review, Ref. \cite{Spa04}). The investigated
systems was subjected to the action of two forces: additive white
noise and regular force. The regular force was fixed or periodical
in time, so that metastable state appeared for a half of period
and unstable state appeared for an other half. The noise enhanced
stability effect implies that, under the action of additive noise,
a system remains in the metastable state for a longer time then in
the deterministic case, and the escape time has a maximum as a
function of noise intensity. We can lengthen or shorten the mean
lifetime of the metastable state of our physical system, by acting
on the white noise intensity. The noise-induced stabilization of
one-dimensional maps \cite{Wac98,Wac99}, the noise-induced
stability in fluctuating bistable potentials \cite{Mie00}, the
noise induced slowing down in a periodical potential
\cite{Dan99,Mah97,Mah98}, the noise induced order in
one-dimensional map of the Belousov-Zhabotinsky reaction
\cite{Mat83,Yos03}, and the transient properties of a bistable
kinetic system driven by two correlated noises \cite{Xie03}, are
akin to the NES phenomenon. Moreover, this effect is at the basis
of resonant trapping \cite{Apo97}. Even though the previous
theoretical papers analyzed NES phenomenon in systems with fixed
or periodically driving metastable and unstable states, the model
of randomly switching metastable state is more realistic in many
cases, e.~g. when we describe the generation process of the
carrier traps in semiconductors. Despite its experimental
importance, the theory of fluctuating barrier crossing is not well
developed for arbitrary noise intensity. In the present paper we
obtain and study analytically the NES effect in a system described
by a potential, which randomly switches between metastable and
unstable configurations. We define the lifetime of metastable
state as the mean first passage time (MFPT) and obtain the
conditions when it can grow with white noise intensity. The mean
first passage time behaviour as a function of the mean flipping
rate of the potential for unstable and metastable initial
configurations is also analyzed. We observe the resonant
activation phenomenon for initial metastable configuration of the
potential profile, and monotonic behaviour for initial unstable
configuration. We can describe therefore, with the same
theoretical approach, two noise-induced phenomena
\cite{Man96,Agu01,Spa04,Doe92,Mar96,ManSpa00}.
\section{The model}
\label{sec:1}

\begin{figure}
\begin{center}
\resizebox{0.84 \columnwidth}{!} {\includegraphics{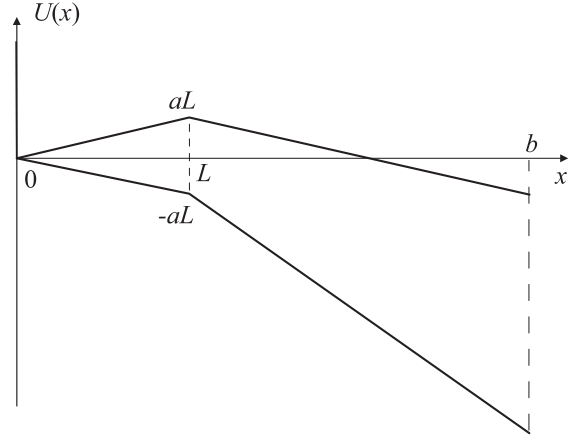}}
\end{center}
\caption{Two configurations of fluctuating potential $U(x)$,
corresponding to the metastable and unstable states.}
\label{fig:1}
\end{figure}

We consider one-dimensional overdamped Brownian motion in a randomly
switching potential profile $U(x)$
\begin{equation}
\frac{dx}{dt}=f\left( x\right) - a\eta \left( t\right) +\xi \left(
t\right) , \label{Lang}
\end{equation}
where $\xi (t)$ is the white Gaussian noise with zero mean and
$\left\langle \xi (t)\xi (t+\tau )\right\rangle =2D\delta (\tau
)$, $\eta (t)$ is a Markovian dichotomous process, which takes the
values $\pm 1$ with mean flipping rate $\nu $, and $f(x) = -
dU(x)/dx$. We investigate the mean first passage time with the
reflecting boundary at the point $x=0$ and the absorbing boundary
at the point $x=b$ ($b>0$). The calculation technique for the MFPT
of non-Markovian process $x(t)$ without white noise ($\xi(t)=0$)
has been originally developed in Ref. \cite{Han85} and, then,
generalized by various authors (see, for example Refs.
\cite{Mas86,Rod86,Doe87}). The exact equations for mean first
passage times, which take into account both the dichotomous and
the white noise terms in Eq.~(\ref{Lang}), were derived in Ref.
\cite{Bal88}, where the authors solve the delicate problem to
construct correct conditions at the reflecting boundary, since all
previous works dealt with absorbing boundaries only. Nevertheless
the cited boundary conditions (3.7a) and (3.7b) in Ref.
\cite{Bal88} are only valid for the special unlikely case of
immediate reflection. For our purposes we use the same conditions
at the reflecting boundary as in Ref. \cite{Zur93}. Thus from the
backward Fokker-Planck equation we obtain the coupled differential
equations for the MFPTs in our system (\ref{Lang}) (see Ref.
\cite{Bal88} and Appendix A)
\begin{eqnarray}
D T^{\prime \prime}_{+} + \left[ f\left( x\right) - a\right]
T^{\prime}_{+} +
\nu \left( T_{-}-T_{+} \right) &=& -1,  \nonumber \\
D T^{\prime \prime}_{-} + \left[ f\left( x\right) + a\right]
T^{\prime}_{-} + \nu \left( T_{+}-T_{-} \right) &=& -1.
\label{Hang}
\end{eqnarray}
Here $T_{+}(x)$ and $T_{-}(x)$ are respectively the mean first
passage times of the boundary $x=b$ for initial values $\eta
(0)=+1$ and $\eta (0)=-1$, with the Brownian particle starting
from the initial point $x$ ($0<x<b$). Our conditions at the
reflecting boundary $x=0$ and the absorbing boundary $x=b$ are
(see Appendix B)
\begin{equation}
T^{\prime}_{\pm }(0) =0, \qquad T_{\pm }\left( b\right) =0.
\label{cond-1}
\end{equation}
Let us introduce for convenience two new auxiliary functions
\begin{equation}
T=\frac{T_{-}+T_{+}}{2}, \qquad \Theta =\frac{T_{-}-T_{+}}{2},
\label{newT}
\end{equation}
and rewrite Eqs.~(\ref{Hang}) in more simple form
\begin{eqnarray}
DT^{\prime \prime}+ f\left( x\right) T^{\prime}+a\Theta ^{\prime
}&=&-1,
\nonumber \\
D\Theta^{\prime \prime}+ f\left( x\right) \Theta
^{\prime}+aT^{\prime}-2\nu \Theta &=&0.
 \label{twoT}
\end{eqnarray}
The boundary conditions for the functions $T(x)$ and $\Theta(x)$
follow from Eqs.~(\ref{cond-1}) and (\ref{newT})
\begin{equation}
T^{\prime}\left( 0\right) =\Theta ^{\prime}\left( 0\right)
=0,\qquad T\left( b\right) =\Theta \left( b\right) =0.
\label{cond-3}
\end{equation}
Equations~(\ref{twoT}) with the boundary conditions (\ref{cond-3})
were solved in Ref. \cite{Zur93} for a model with constant force
\begin{equation}
f(x)=const., \qquad 0<x<b. \label{forceDoe}
\end{equation}
In the present paper we consider more general model with step-wise
force (or piece-wise potential $U(x)$) as
\begin{equation}
f(x)= - \frac{dU(x)}{dx} = k\cdot \theta(x-L), \qquad 0<L<b,
\label{force}
\end{equation}
where $\theta(x)$ is the step function and $k$ is a constant. As
we can see from Fig.~\ref{fig:1}, if $0<a<k$ we have a metastable
state for $\eta (t)=+1$ and an unstable state for $\eta (t)= -1$.

In particular cases when $L\to 0$ or $L\to b$ the force
(\ref{force}) coincides with (\ref{forceDoe}). After removing
$T(x)$ from Eq.~(\ref{twoT}) we obtain third-order linear
differential equation for the variable $\Theta (x)$
\begin{equation}
\Theta ^{\prime \prime \prime}+\frac{2f\left( x\right) }{D}\Theta
^{\prime \prime}+ F(x) \Theta ^{\prime}-\frac{2\nu f\left(
x\right) }{D^{2}}\Theta =\frac{a}{D^{2}}, \label{theta}
\end{equation}
where
\begin{equation}
F(x) = \frac{f^{2}\left( x\right) }{D^{2}}+\frac{f^{'}\left(
x\right) }{D}-\gamma ^{2}, \quad
 \gamma
=\sqrt{\frac{a^{2}}{D^{2}}+\frac{2\nu }{D}}. \label{gamma}
\end{equation}
We will consider the mean first passage times $T_{\pm}(0)$ with
the starting position of Brownian particle at $x=0$. In the
absence of switchings and white noise the dynamical escape time
$T_{+}(0)$ for the initial metastable state is equal to $+\infty$
and the dynamical escape time $T_{-}(0)$ for the initial unstable
state equals
\begin{equation}
T_{-}(0)=\frac{L}{a}+\frac{b-L}{k+a}. \nonumber
\end{equation}
We solve Eqs.~(\ref{twoT}) and (\ref{theta}) for regions $0<x<L$
and $L<x<b$ separately. First of all we find the solutions in the
interval $\left( L,b\right)$
\begin{eqnarray}
\Theta \left( x\right) &=& \sum_{i=1}^{3} c_{i}e^{\lambda
_{i}\left( x-L\right)}
-\frac{a}{2\nu k},  \nonumber \\
T\left( x\right) &=& c_{4}-\frac{x-L}{k}-a\sum_{i=1}^{3}
\frac{c_{i}\left[ e^{\lambda _{i}\left( x-L\right)}-1\right]
}{k+\lambda _{i}D}, \label{Ttheta-2}
\end{eqnarray}
where $c_{i}$ $(i=1\div4)$ are unknown constants and $\lambda
_{1}$, $\lambda _{2}$, $\lambda _{3}$ are the roots of the
following cubic equation
\begin{equation}
\lambda \left( \lambda D+k\right) ^{2}-\Gamma ^{2}\lambda - 2\nu
k=0, \label{roots}
\end{equation}
where $\Gamma=\gamma D$. Using graphical representation it can be
easily shown that algebraic Eq.~(\ref {roots}) has three real
roots: one positive and two negative ones.

Taking into account the conditions (\ref{cond-3}) at the
reflecting boundary $x=0$ we obtain for region $(0,L)$
\begin{eqnarray}
\Theta \left( x\right) &=&c_{5}\left( 1+\frac{2\nu D}{a^{2}}\cosh
{\gamma x} \right) +\frac{a}{\Gamma ^{3}}\left( D\sinh {\gamma
x}-\Gamma x\right) ,
\nonumber \\
T\left( x\right) &=&-\frac{\nu x^{2}}{\Gamma
^{2}}-\frac{a^{2}D}{\Gamma
^{4}}\left( \cosh {\gamma x}-1\right) \nonumber \\
&& - \ c_{5}\frac{2\nu }{\Gamma a}\left( D\sinh {\gamma x} -\Gamma
x\right) +c_{6}. \label{Ttheta-1}
\end{eqnarray}
From Eqs.~(\ref{newT}) and (\ref{Ttheta-1}) we get finally
\begin{equation}
T_{\pm}\left( 0\right) =T\left( 0\right) \mp \Theta \left(
0\right) =c_{6} \mp c_{5}\frac{ \Gamma ^{2}}{a^{2}}. \label{T-}
\end{equation}
The six unknown constants $c_{i}$ $(i=1\div6)$ can be determined
from the conditions (\ref{cond-3}) at the absorbing boundary $x = b$
and the continuity conditions of the functions $\Theta (x)$,
$\Theta ^{\prime}(x)$, $T(x)$, $T^{\prime }(x)$ at the point
$x=L$. Thus, we obtain the following system of algebraic linear
equations for the constants $c_i$
\begin{eqnarray}
&&\sum_{i=1}^{3}c_{i}e^{\mu _{i}} = \frac{a}{2\nu k},
\nonumber\\
&&\sum_{i=1}^{3} \frac{c_{i}a\left( e^{\mu _{i}}-1\right)
}{k+\lambda _{i}D} =
c_4-\frac{b-L}{k},\nonumber\\
&&\sum_{i=1}^{3}c_{i} = c_{5}\left( 1+\frac{2\nu D}{a^{2}}\cosh
{\gamma L}
\right)\nonumber \\
&& +\frac{a}{\Gamma ^{3}}\left( D\sinh {\gamma L}-\Gamma L\right)+
\frac{a}{2\nu k},   \nonumber\\
&&\sum_{i=1}^{3}c_{i}\lambda _{i} = c_{5}\frac{2\nu
\Gamma}{a^{2}}\sinh {\gamma L}+\frac{a}{\Gamma^{2}}\left( \cosh {
\gamma L}-1\right) ,  \nonumber \\
&&\sum_{i=1}^{3} \frac{c_{i}\lambda _{i}a}{k+\lambda _{i}D} =
\frac{2\nu L}{\Gamma ^{2}}+\frac{a^{2}}{\Gamma
^{3}} \sinh {\gamma L} \nonumber \\
&& + \frac{2\nu }{a}c_{5}\left( \cosh {\gamma L}-1\right)
-\frac{1}{k} ,\nonumber \\
&&c_{4} = c_{6} - \frac{\nu L^{2}}{\Gamma
^{2}} - \frac{a^{2}D}{\Gamma^{4}} \left( \cosh {\gamma L}-1\right) \nonumber \\
&& - c_{5}\frac{2\nu }{\Gamma a}\left( D\sinh {\gamma L}-\Gamma
L\right),
\label{third}
\end{eqnarray}
where $\mu_i = \lambda_i (b - L)$. The equation (\ref{T-}) with
algebraic system (\ref{third}) is the exact solution of
Eqs.~(\ref{Hang}) for the force (\ref{force}). In the limits
$L\to0$ or $L\to b$, the expression (\ref{T-}) coincides with that
obtained in Ref.~\cite{Zur93} for the force (\ref{forceDoe}).

\section{NES phenomenon conditions}
\label{sec:2}

Our aim is to investigate the noise enhanced stability effect. In
other words we are interested in the situation, when the escape
time grows with noise. We expect to find this growing in the limit
of small intensity of the white noise. Hereafter we write the mean
first passage times $T_{\pm}(0)$ as $T_{\pm}(\nu,D)$ to point out
the dependence on the parameters $\nu$ and $D$ of two noise
sources. On cumbersome rearrangements (see Appendix C), from
Eqs.~(\ref{T-}) and (\ref{third}) we obtain in the limit
$D\rightarrow 0$
\begin{equation}
T_{-}(\nu,D) \simeq T_{-}(\nu,0) + \frac{D}{a^{2}}G(q,\omega,s) +
o\left( D\right), \label{T-approx}
\end{equation}
where
\begin{equation}
T_{-}(\nu,0) =\frac{2L}{a}+\frac{\nu L^{2}}{a^{2}}+\frac{b-L}{k}
-\frac{q\left( 1-q\right) }{2\nu }\left( 1-e^{-s}\right)
\label{zero}
\end{equation}
is the MFPT in the absence of white noise, and
\begin{eqnarray}
G(q,\omega,s) &=& \frac{q^{3}\left( 1+q^{2}\right) se^{-s}}{\left(
1+q\right) \left( 1-q^{2}\right)
}-\frac{5+q-5q^{2}-5q^{3}}{2\left( 1+q\right) \left(
1-q^{2}\right) } \nonumber \\
&+& \frac{q\left(1+q-5q^{2}-3q^{3}-2q^{4}\right) }{2\left(
1+q\right) \left( 1-q^{2}\right) }\left( 1-e^{-s}\right) \nonumber
\\
&+& \frac{2\omega \left( 3q^{2}+q-3\right) }{q\left(
1-q^{2}\right) }- \frac{2\omega ^{2}}{q^{2}}. \label{main}
\end{eqnarray}
Here $q$, $\omega $ and $s$ are the dimensionless parameters
\begin{equation}
q = \frac{a}{k},\quad \omega = \frac{\nu L}{k},\quad s =
\frac{2\omega }{1-q^{2}} \left( \frac{b}{L}-1\right) .
\end{equation}
As one would expect, the MFPT (\ref {zero}) coincides with the nonlinear
relaxation time (NLRT) for the same system (see Eq.~(32) in
Ref.~\cite{Dub04}) because in the absence of diffusion $(D=0)$,
after crossing the point $x=b$ Brownian particles cannot come back
in the interval $(0,b)$ again.

\begin{figure}
\begin{center}
\resizebox{0.84\columnwidth}{!}{\includegraphics{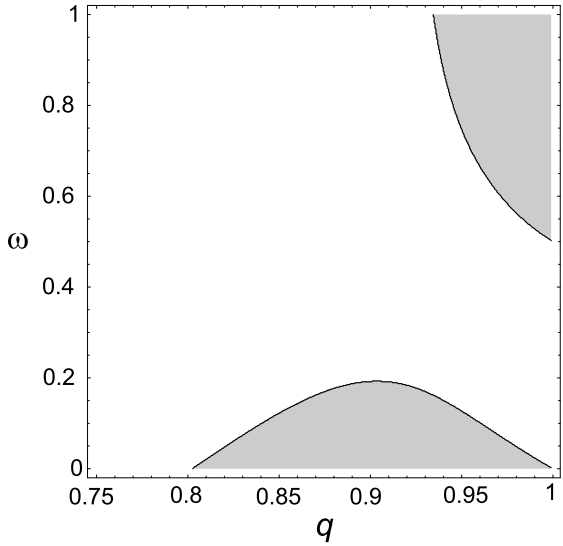}}
\end{center}
\caption{The shaded areas are the regions of the plane ($q,
\omega$) where the NES effect takes place. The parameter is
$b/L=1.5$.} \label{fig:2}
\end{figure}

The sign of the term $G(q,\omega,s)$ in Eq.~(\ref{main}) allows to
obtain the conditions to observe the NES effect in the system
investigated. We can write these conditions as
\begin{equation}
G(q,\omega,s)>0. \label{NES}
\end{equation}
Let us analyse the structure of NES phenomenon region on the plane
$(q,\omega )$. In the case of very slow switchings $\nu
\rightarrow 0$ $(\omega \rightarrow 0,\;s\rightarrow 0)$
Eq.~(\ref{NES}) takes the form
$$
\omega < \frac{q\left( 1-q\right) \left( 5q^{3}+5q^{2}-q-5\right)
}{2\left[ 6-2q-q^{2}\left( b/L+11\right) +2q^{3}+3q^{4}\left(
b/L+1\right) \right] },
$$
\begin{equation}
q > 0,8024. \label{regions}
\end{equation}
In the case of $q\simeq 1$ we obtain from Eq.~(\ref{NES})
\begin{equation}
\omega < \frac{x_0 \left( 1-q\right) }{b/L-1},\quad
\frac{1}{2}+\frac{5}{2}\left( 1-q\right) < \omega <
\frac{1}{2(1-q)},
\end{equation}
where $x_0$ is the positive root of algebraic equation
$$
e^x=x+2
$$
approximately equals $x_0 \simeq 1.1463$. The two shaded areas on
the plane $(q,\omega )$, where the NES phenomenon takes place, are
shown in Fig.~\ref{fig:2}. Both noise enhanced stability areas
connect at the parameter $b/L<1.2655$.

Now we may conclude that the NES effect occurs at the values of
$q$ near $1$, i.e. at very small steepness $k-a=k(1-q)$ of the
reverse potential barrier for the metastable state (see
Fig.~\ref{fig:1}). Only in such a situation, Brownian particles
can move back into potential well in the interval $(0,L)$ from the
region $L<x<b$, with a low noise intensity. In Ref. \cite{Agu01}
we obtained the parameter region of NES effect for periodical
driven metastable state. This region is larger than that obtained
here for randomly switching potential. In our opinion, this
narrowing of NES phenomenon area is attributable to the
uncertainty of the first switching time. Moreover, because of the
exponential probability distribution of the time interval between
jumps of the dichotomous noise, which randomly switches our
metastable state, the average escape time has a large variance and
thus the analogy between the two cases falls.

It must be emphasized that noise enhanced stability regions of the
mean first passage time are inside that obtained for the nonlinear
relaxation time with the same parameter $b/L$ (compare
Eq.~(\ref{main}) with Eq.~(31) in Ref.~\cite{Dub04}) because the
nonlinear relaxation time takes into account a repeated reentries
of Brownian particles in the interval $(0,b)$. The NES effect
disappears when the absorbing boundary is placed at the point
$x=L$. In fact by putting $b\rightarrow L$, $k\rightarrow+\infty $
($q\rightarrow 0$, $s\rightarrow 0$) in Eq.~(\ref{T-approx}) we
obtain
\begin{equation}
T_{-}(0)\simeq \frac{2L}{a}+\frac{\nu
L^{2}}{a^{2}}-\frac{D}{a^{2}}\left( \frac{5}{2}+\frac{6\nu
L}{a}+\frac{2\nu ^{2}L^{2}}{a^{2}}\right) .
\end{equation}

\begin{figure}
\begin{center}
\resizebox{0.84\columnwidth}{!}{\includegraphics{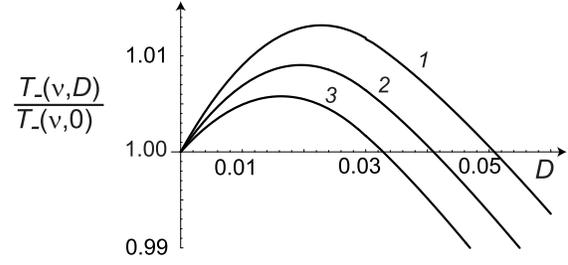}}
\end{center}
\caption{The normalized MFPT vs the white noise intensity $D$, for
three values of the dimensionless mean flipping rate $\omega =
(\nu L)/k$: 0.01 (curve 1), 0.04 (curve 2), 0.07 (curve 3) (lower
area in Fig.~\ref{fig:2}). Parameters are $a=0.9$, $k=1$, $L=1$,
and $b=1.5$.} \label{fig:3}
\end{figure}

\begin{figure}
\begin{center}
\resizebox{0.84\columnwidth}{!}{\includegraphics{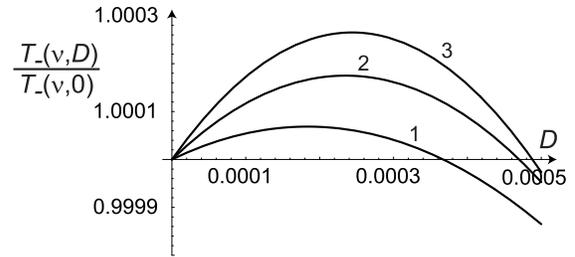}}
\end{center}
\caption{The normalized MFPT vs the white noise intensity $D$, for
three values of the dimensionless mean flipping rate $\omega =
(\nu L)/k$: 0.7 (curve 1), 0.8 (curve 2), 0.9 (curve 3) (upper
area in Fig.~\ref{fig:2}). Parameters are $a=0.97$, $k=1$, $L=1$,
and $b=1.5$.} \label{fig:4}
\end{figure}

The behaviours of the MFPTs $T_-(\nu,D)$ for the initial unstable
configuration, normalized to the value obtained in the absence of
white noise $T_-(\nu,0)$, vs the white noise intensity are shown
in Figs.~\ref{fig:3} and \ref{fig:4} for the two NES areas
represented in Fig.~\ref{fig:2}.
As we can see the effect decreases, for fixed $q$, when $\omega$
decreases in the upper area of Fig.~\ref{fig:2}, and when $\omega$
increases in the lower area of Fig.~\ref{fig:2}. At the same time, the NES
phenomenon for the upper area in Fig.~\ref{fig:2} is very small
(see Fig.~\ref{fig:4}).

\begin{figure}
\begin{center}
\resizebox{0.84\columnwidth}{!}{\includegraphics{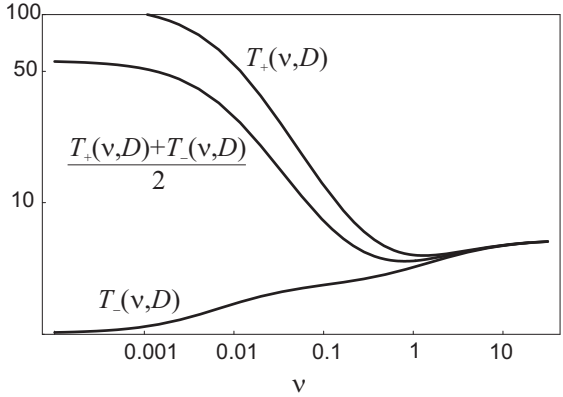}}
\end{center}
\caption{Logarithmic plot of the mean first passage times:
$T_{+}(\nu,D)$ (initial metastable configuration), $T_{-}(\nu,D)$
(initial unstable configuration), $\bar T(\nu,D)$ (arithmetic
average), as a function the mean flipping rate $\nu$. Here $D =
0.1$, $a = 0.5$ and the other parameters are the same as in
Figs.~\ref{fig:3} and \ref{fig:4}.} \label{fig:5}
\end{figure}

Now we analyse the behaviour of the escape times $T_-(\nu,D)$ and
$T_+(\nu,D)$ as a function of the mean flipping rate $\nu$. From
Eqs.~(\ref{T-}) and (\ref{third}) we obtain these behaviours,
which are shown in Fig.~\ref{fig:5} for a fixed value of the white
noise intensity $D = 0.1$.

We find the resonant activation phenomenon \cite{Doe92,ManSpa00}
for $T_{+}(\nu,D)$ and monotonic behaviour for $T_{-}(\nu,D)$. In
the same Fig.~\ref{fig:5} we report the arithmetic average of
these quantities: $\bar T(\nu,D) = \left[ T_{+}(\nu,D) +
T_{-}(\nu,D)\right] /2$, which is the average escape time analyzed
by Doering and Gadoua in Ref. \cite{Doe92}. Starting from the
metastable configuration, the mean first passage time
$T_{+}(\nu,D)$ decreases with the mean flipping rate, because the
potential fluctuations induce crossing events of the potential
barrier. On the other hand, starting from the unstable
configuration, the average escape time $T_{-}(\nu,D)$ increases
monotonically because the potential fluctuations stabilise the
initial unstable state. For very fast potential fluctuations,
Brownian particles "forget" the initial configuration of the
potential and all the mean first passage times ($T_{+}(\nu,D)$,
$T_{-}(\nu,D)$, and $\bar T(\nu,D)$) join in the same asymptotic
value: the average escape time required to cross the average
barrier \cite{Doe92,ManSpa00}.

\section{Conclusions}
\label{sec:3}

We have investigated the noise enhanced stability (NES) phenomenon
in one-dimensional system with Gaussian additive white noise and a
potential randomly switched by a Markovian dichotomous process. We
have derived general equations to calculate the mean first passage
times with one reflecting boundary and one absorbing boundary. For
piece-wise linear potential we obtain the exact solution for MFPTs
as a function of arbitrary white noise intensity and arbitrary
mean flipping rate of the potential barrier. By analysing the
derived equations we obtain analytically the region of the NES
phenomenon occurrence. We find that this effect can be observed
only at very flattened sink beyond the potential barrier, i.e. at
the real absence of the reverse potential barrier in the
metastable state. Moreover we note that the noise enhanced
stability phenomenon is related to rare events, that is when some
particles are trapped in the metastable state. Therefore to better
understand this effect it should be interesting to analyse the
probability distribution of the escape time. This will be done in
a forthcoming paper. By investigating the behaviour of the mean
first passage times as a function of the mean flipping rate of the
potential we find the resonant activation phenomenon for initial
metastable configuration and monotonic behaviour for initial
unstable configuration. With our theoretical approach we are able
to describe, therefore, two different noise-induced effects. Our
exact analytical results, concerning the enhancement of stability
of metastable states for one-dimensional systems, can be a good
starting point to extend and improve the same theoretical
apparatus to more complex systems with fluctuating potential
barriers.

\begin{acknowledgement}
This work has been supported by INTAS Grant 2001-0450, MIUR, INFM,
Russian Foundation for Basic Research (project 02-02-17517),
Federal Program "Scientific Schools of Russia" (project
1729.2003.2), and by Scientific Program "Universities of Russia"
(project 01.01.020).
\end{acknowledgement}

\section{Appendix A}

Let us consider a stochastic Langevin equation with two random
forces
\begin{equation}
\dot{x}=f\left(  x\right)  +g\left(  x\right)  \eta\left( t\right)
+h\left( x\right)  \xi\left(  t\right)  , \label{Lang-2}
\end{equation}
where $\xi\left(  t\right)  $ is the white Gaussian noise with
zero mean and correlation function $\left\langle \xi\left(
t\right) \xi\left( t+\tau\right)  \right\rangle =2D\delta\left(
\tau\right)$, $\eta\left( t\right)$ is a Markovian dichotomous
noise with flippings mean rate $\nu$ and values $\pm1$, $f\left(
x\right) $, $g\left( x\right)  $, and $h\left(  x\right)  $ are
arbitrary functions. To derive the closed equation for the joint
probability density of the random processes $x(t)$ and $\eta (t)$
we start from the following expression

\begin{equation}
W\left( x,y,t\right) =\left\langle\delta \left( x-x(t)\right)
\delta \left( y-\eta(t)\right) \right\rangle  \label{known}
\end{equation}
and then apply well-known method based on functional correlation
formulae (see, for example, Refs. \cite{Luc92,Dub04}).

Upon differentiation of Eq.~(\ref{known}) on $t$, we obtain
\begin{eqnarray}
\frac{\partial W}{\partial t}&=&-\frac{\partial}{\partial
x}\left\langle \dot {x}\left(  t\right)  \delta\left(  x-x\left(
t\right) \right) \delta \left( y-\eta(t)\right) \right\rangle \nonumber \\
&+& \left\langle\delta (x-x(t))\frac{\partial}{\partial t}\delta
(y-\eta(t))\right\rangle . \label{W-1}
\end{eqnarray}
Substituting $\dot{x}\left( t\right) $ from Eq.~(\ref{Lang-2}) and
taking into account the definition (\ref{known}), we can rewrite
Eq.~(\ref{W-1}) as
\begin{eqnarray}
&&\frac{\partial W}{\partial t} = -\frac{\partial}{\partial
x}\left[ f(x)+yg(x)\right] W\nonumber \\
&&-\frac{\partial}{\partial x}h(x)\left\langle \xi\left( t\right)
\delta\left( x-x\left( t\right) \right) \delta \left(
y-\eta(t)\right) \right\rangle \nonumber \\
&&+ \left\langle\delta (x-x(t))\frac{\partial}{\partial t}\delta
(y-\eta(t))\right\rangle . \label{W-2}
\end{eqnarray}
To split the functional average in Eq.~(\ref{W-2}) we use
Furutsu-Novikov's formula for the white Gaussian noise $\xi\left(
t\right) $ \cite{Fur63,Nov65}
\begin{eqnarray}
\left\langle \xi\left( t\right) F_{t}\left[ \xi\right]
\right\rangle =D\left\langle \frac{\delta F_{t}\left[ \xi\right]
}{\delta\xi\left( t\right) }\right\rangle , \label{Fur-Nov}
\end{eqnarray}
where $F_{t}\left[ \xi \right] $ is an arbitrary functional
depending on the history of random process $\xi\left( t\right) $
($0 \leq \tau\leq t$). Replacing $F_{t}\left[ \xi \right] $ with
$\delta \left(  x-x\left(t\right) \right) \delta \left( y-\eta
\left(t\right) \right) $ in Eq.~(\ref{Fur-Nov}) and taking into
account that, in accordance with Eq.~(\ref{Lang-2}), $\delta
x\left( t\right) /\delta\xi\left(t\right) = h\left( x(t)\right) $,
we obtain
\begin{equation}
\left\langle \xi \left( t\right) \delta \left( x-x\left( t\right)
\right) \delta \left( y-\eta \left( t\right) \right) \right\rangle
=-D\frac{\partial }{\partial x}h\left( x\right) W. \label{ave}
\end{equation}

Further let us rearrange the last part of second average in
Eq.~(\ref{W-2}). Using the translation operator and taking into
account that for dichotomous noise with the values $\pm 1$:
$\eta^{2k}\left( t\right) = 1$, $\eta^{2k+1}\left( t\right) = \eta
\left(  t\right) $ ($k=1,2,\ldots $), we arrive at
\begin{eqnarray}
&&\frac{\partial}{\partial t}\delta \left( y-\eta(t)\right)
=\frac{\partial}{\partial t}\exp \left(
-\eta(t)\frac{d}{dy}\right) \delta \left( y\right)
=\nonumber \\
&&-\frac{\partial}{\partial t}\sinh \left(
\eta(t)\frac{d}{dy}\right) \delta \left( y\right) = -
\dot{\eta}\left(  t\right) \sinh \left( \frac{d}{dy}\right) \delta
(y).\qquad \label{W-3}
\end{eqnarray}
Substituting Eqs.~(\ref{ave}) and (\ref{W-3}) in Eq.~(\ref{W-2}),
we obtain
\begin{eqnarray}
\frac{\partial W}{\partial t} &=& -\frac{\partial}{\partial
x}\left[ f(x)+yg(x)\right] W+D\frac{\partial}{\partial x}h(x)
\frac{\partial }{\partial x}h\left( x\right) W\nonumber \\
&-& \left\langle \dot{\eta}\left(  t\right) \delta (x-x(t))\sinh
\left( \frac{d}{dy}\right) \delta (y)\right\rangle . \label{W-4}
\end{eqnarray}

Now we use the Shapiro-Loginov's formula for a Markovian
dichotomous noise $\eta \left( t\right)$ \cite{Sha78}, to express
the functional average in Eq.~(\ref{W-4}) in terms of the joint
probability distribution
\begin{equation}
\left\langle \dot{\eta}\left(  t\right)  R_{t}\left[ \eta\right]
\right\rangle =-2\nu\left\langle \eta\left(  t\right) R_{t}\left[
\eta\right]  \right\rangle ,
\label{Sha-Log}
\end{equation}
where $R_{t}\left[ \eta \right] $ is an arbitrary functional
depending on the history of random process $\eta\left( \tau\right)
$ ($0 \leq \tau\leq t$). As a result, for the functional average
in Eq.~(\ref{W-4}) we have
\begin{eqnarray}
&&\left\langle \dot{\eta}\left(  t\right) \delta (x-x(t))\sinh
\left( \frac{d}{dy}\right) \delta (y)\right\rangle =\nonumber \\
&&-2\nu \left\langle \delta (x-x(t))\sinh \left( \eta\left(
t\right) \frac{d}{dy}\right) \delta (y)\right\rangle =\nonumber
\\
&&- \nu \left\langle \delta (x-x(t))\left[ \delta (y+\eta (t))-
\delta (y-\eta (t))\right] \right\rangle .\qquad \label{W-5}
\end{eqnarray}

Substituting Eq.~(\ref{W-5}) in Eq.~(\ref{W-4}) and taking into
account Eq.~(\ref{known}) we obtain finally the following forward
Kolmogorov's equation for the joint probability distribution
\begin{eqnarray}
\frac{\partial W}{\partial t} &=& -\frac{\partial}{\partial
x}\left[ f\left(  x\right)  +yg\left(  x\right)  \right] W+D\left[
\frac{\partial }{\partial x}h\left( x\right)\right]^2W\nonumber \\
&+& \nu\left[ W(x,-y,t)-W(x,y,t)\right] . \label{Direct}
\end{eqnarray}

According to Eq.~(\ref{Direct}) we can reconstruct the backward
Kolmogorov's equation for the conditional probability distribution
$W(x,y,t\left\vert x_{0},y_{0},t_{0}\right. )$ of the Markovian
vector-process $\left\{ x\left(  t\right) ,\eta\left( t\right)
\right\}$
\begin{eqnarray}
&&-\frac{\partial W}{\partial t_{0}}=\left[  f\left( x_{0}\right)
+y_{0}g\left(  x_{0}\right)  \right]  \frac{\partial W}{\partial
x_{0} }+D\left[ h\left(  x_{0}\right) \frac{\partial}{\partial
x_{0}}\right]^2W\nonumber \\
&&+ \nu\left[  W(x,y,t\left\vert x_{0},-y_{0},t_{0}\right.
)-W(x,y,t\left\vert x_{0},y_{0},t_{0}\right.  )\right]. \qquad
\label{Backward}
\end{eqnarray}
Since the conditional probability density depends only on the
difference $\tau=t-t_{0}$, Eq.~(\ref{Backward}) can be rewritten
as
\begin{eqnarray}
&&\frac{\partial W}{\partial\tau}=\left[  f\left( x_{0}\right)
+y_{0}g\left(  x_{0}\right)  \right]  \frac{\partial W}{\partial
x_{0} }+D\left[ h\left( x_{0}\right) \frac{\partial}{\partial
x_{0}}\right]^2W\nonumber \\
&&+ \nu\left[  W(x,y,\tau\left\vert x_{0},-y_{0},0\right.
)-W(x,y,\tau\left\vert x_{0},y_{0},0\right. )\right]. \qquad
\label{New}
\end{eqnarray}

Let us introduce the random first passage time $\theta$. The
probability $\Pr\left\{\theta>\tau\right\}\equiv P\left(
x_{0},y_{0},\tau\right)$ that random process $x\left( t\right) $
remains between the absorbing boundaries $x=L_{1}$ and $x=L_{2}$
during the time interval $\left( 0,\tau\right) $ is
\begin{equation}
P\left(  x_{0},y_{0},\tau\right)
=\int_{L_{1}}^{L_{2}}dx\int_{-\infty}^{\infty}W(x,y,\tau\left\vert
x_{0},y_{0},0\right. )dy. \label{prob}
\end{equation}
Taking into account Eq.~(\ref{prob}) and integrating
Eq.~(\ref{New}) on $x$ and $y$ we find
\begin{eqnarray}
&&\frac{\partial P}{\partial\tau} = \left[  f\left(  x_{0}\right)
+y_{0}g\left( x_{0}\right) \right] \frac{\partial P}{\partial
x_{0}}+D\left[ h\left( x_{0}\right) \frac{\partial}{\partial
x_{0}}\right]^2P\nonumber \\
&&+ \nu\left[  P(x_{0},-y_{0},\tau )-P(x_{0},y_{0},\tau)\right].
\label{Prel-2}
\end{eqnarray}
To obtain the probability density of first passage time we should
differentiate the probability $\Pr\left\{  \theta>\tau\right\}$ on
$\tau$
$$
w\left(  x_{0},y_{0},\tau\right)  =\frac{\partial\Pr\left\{ \theta
<\tau\right\}  }{\partial\tau} = -\frac{\partial
P(x_{0},y_{0},\tau )}{\partial\tau}.
$$
As a result, we have the same Eq.~(\ref{Prel-2}) for $w\left(
x_{0},y_{0},\tau\right)$
\begin{eqnarray}
&&\frac{\partial w}{\partial\tau} = \left[  f\left(  x_{0}\right)
+y_{0}g\left( x_{0}\right) \right] \frac{\partial w}{\partial
x_{0}}+D\left[ h\left( x_{0}\right) \frac{\partial}{\partial
x_{0}}\right]^2w\nonumber \\
&&+ \nu\left[  w(x_{0},-y_{0},\tau )-w(x_{0},y_{0},\tau)\right].
\label{Prel-3}
\end{eqnarray}

For the mean first passage time
\begin{equation}
\vartheta\left(  x_{0},y_{0}\right)  =\int_{0}^{\infty}\tau
w(x_{0},y_{0} ,\tau)d\tau
\label{warthe}
\end{equation}

we obtain from Eq.~(\ref{Prel-3}) the following differential
equation
\begin{eqnarray}
&&D\left[ h\left(  x_{0}\right) \frac{\partial}{\partial
x_{0}}\right]^2\vartheta+\left[ f\left( x_{0}\right) +y_{0}g\left(
x_{0}\right) \right] \frac{\partial\vartheta}{\partial
x_{0}}\nonumber \\
&&+\nu\left[ \vartheta(x_{0},-y_{0})-\vartheta(x_{0},y_{0})\right]
=-1. \label{Eq-mean}
\end{eqnarray}
Since $y_{0}=\pm1$, Eq.~(\ref{Eq-mean}) is equivalent to the
following set of equations for times $T_{+}\left( x_{0}\right)
\equiv\vartheta(x_{0},+1) $ and $T_{-}\left( x_{0}\right)
\equiv\vartheta(x_{0},-1)$
\begin{eqnarray}
\hat{L}_{x_0}T_{+} + \left[ f\left( x_{0}\right) +g\left(
x_{0}\right) \right]
\frac{dT_{+}}{dx_{0}}+\nu\left(T_{-}-T_{+}\right)
&=&-1,\qquad \nonumber\\
\hat{L}_{x_0}T_{-}+\left[ f\left( x_{0}\right)  -g\left(
x_{0}\right)  \right]
\frac{dT_{-}}{dx_{0}}+\nu\left(T_{+}-T_{-}\right) &=& -1, \quad
\label{New-Han}
\end{eqnarray}
where we introduced the operator
\begin{equation}
\hat{L}_{x_0} = D \left[ h(x_0)\frac{d}{dx_{0}}\right]^2.
\label{operator-L}
\end{equation}

Equations (\ref{New-Han}) are similar to well-known equations for
mean first passage times obtained in Ref. \cite{Bal88}. But
authors used the Ito's interpretation of Langevin equation
(\ref{Lang-2}) and, as a result, obtained different equations. For
$h\left( x\right) =1$, Eqs.~(\ref{New-Han}) coincide with the
equations derived in Ref. \cite{Bal88}. Finally, if we put in
Eqs.~(\ref{New-Han}) and (\ref{operator-L}): $g\left( x\right)
=-a$, $h\left( x\right) =1$, we obtain Eqs.~(\ref{Hang}).

\section{Appendix B}

We explain the origin of conditions (\ref{cond-1}) for the mean
first passage times at a reflecting boundary. For sake of
definiteness, we consider a left reflecting boundary at the point
$A$. Let Brownian particle reaches the reflecting boundary at the
instant $t_{0}-\Delta t_{0}$. For the next small time period
$\Delta t_{0}$, dichotomous noise $\eta (t)$ can switch to the
opposite state with the probability $\beta(\Delta t_{0})$ or
remains in the initial state with the probability $\left(
1-\beta(\Delta t_{0})\right) $. The scenario of Brownian particle
behaviour for this time interval is the following: it transfers to
the point $A+\Delta x_{0}$ ($\Delta x_{0}>0$) with some fixed
probability $\alpha $ or stays at reflecting boundary with the
probability $\left( 1-\alpha \right) $. We denote the probability
$\alpha $ as $\alpha_{-}$ for the case when dichotomous noise
switches and as $\alpha_{+}$ for the opposite case. As a result,
the joint conditional probability distribution reads
\begin{eqnarray}
&&W(x_{0},z_{0},t_{0}\left\vert A,y_{0},t_{0}-\Delta t_{0}\right.
)=\beta (\Delta t_{0})\delta \left( z_{0}+y_{0}\right) \cdot \nonumber \\
&&\left[ \alpha_{-}\delta \left( x_{0}-A-\Delta x_{0}\right)
+\left( 1-\alpha_{-}\right) \delta \left( x_{0}-A\right) \right]
\nonumber \\
&&+\left[ 1-\beta (\Delta t_{0})\right] \delta \left(
z_{0}-y_{0}\right) [\alpha_{+}\delta \left(
x_{0}-A-\Delta x_{0}\right) \nonumber \\
&&+\left( 1-\alpha_{+}\right) \delta \left( x_{0}-A\right) ] .
\label{B-1}
\end{eqnarray}
Substituting Eq.~(\ref{B-1}) in Smoluchowski equation
\begin{eqnarray}
&&W(x,y,t\left\vert A,y_{0},t_{0}-\Delta t_{0}\right.
)=\int\nolimits_{A}^{+\infty}dx_{0}\int\nolimits_{-\infty}^{+\infty}dz_{0}
\nonumber \\
&&W(x_{0},z_{0},t_{0}\left\vert A,y_{0},t_{0}-\Delta t_{0}\right.
)W(x,y,t\left\vert x_{0},z_{0},t_{0}\right. ) \nonumber
\end{eqnarray}
we obtain
\begin{eqnarray}
&&W(x,y,t\left\vert A,y_{0},t_{0}-\Delta t_{0}\right. )\nonumber \\
&&=\alpha_{-}\beta (\Delta t_{0})W(x,y,t\left\vert A+\Delta
x_{0},-y_{0},t_{0}\right.
)\nonumber \\
&&+\alpha_{+}\left[ 1-\beta (\Delta t_{0})\right]
W(x,y,t\left\vert A+\Delta
x_{0},y_{0},t_{0}\right. )\nonumber \\
&&+\beta (\Delta t_{0})\left( 1-\alpha_{-}\right)
W(x,y,t\left\vert
A,-y_{0},t_{0}\right. )\nonumber \\
&&+\left( 1-\alpha_{+}\right) \left[ 1-\beta (\Delta t_{0})\right]
W(x,y,t\left\vert A,y_{0},t_{0}\right. ).\qquad \label{B-2}
\end{eqnarray}
By expanding the conditional probability distributions in
Eq.~(\ref{B-2}) in power series in $\Delta t_{0}$, up to a linear
terms, and in $\Delta x_{0}$, up to a quadratic terms, we find
approximately
\begin{eqnarray}
&&-\Delta t_{0}\frac{\partial W_{+}}{\partial t_{0}}\simeq \beta
(\Delta t_{0})\left( W_{-}-W_{+}\right) \nonumber \\
&&+\alpha_{-}\beta (\Delta t_{0})\Delta x_{0}\left( \left.
\frac{\partial W_{-}}{\partial x_{0}}\right\vert _{A}+\frac{\Delta
x_{0}}{2}\left. \frac{\partial^2W_{-}}{\partial
x_{0}^2}\right\vert _{A}\right) \nonumber \\
&&+\alpha_{+}\Delta x_{0}\left[ 1-\beta (\Delta t_{0})\right]
\left( \left. \frac{\partial W_{+}}{\partial x_{0}}\right\vert
_{A}+\frac{\Delta x_{0}}{2}\left. \frac{\partial^2W_{+}}{\partial
x_{0}^2}\right\vert _{A}\right) ,\qquad \label{B-3}
\end{eqnarray}
where for simplicity we introduce the notations
$W_{+}=W(x,y,t\left\vert A,y_{0},t_{0}\right. )$ and
$W_{-}=W(x,y,t\left\vert A,-y_{0},t_{0}\right. )$.

For Markovian dichotomous noise we have $\beta (\Delta t_{0})=\nu
\Delta t_{0}$. After substituting this value in Eq.~(\ref{B-3})
and dividing by $\alpha_{+}\Delta x_{0}$ both sides of the
equation, we arrive at
\begin{eqnarray}
&&-\frac{\Delta t_{0}}{\alpha_{+}\Delta x_{0}}\frac{\partial
W_{+}}{\partial t_{0}} \simeq \frac{\nu \Delta
t_{0}}{\alpha_{+}\Delta x_{0}}\left(
W_{-}-W_{+}\right) \nonumber \\
&&+\left( 1-\nu \Delta t_{0}\right) \left( \left. \frac{\partial
W_{+}}{\partial x_{0}}\right\vert _{A}+\frac{\Delta
x_{0}}{2}\left. \frac{\partial^2W_{+}}{\partial
x_{0}^2}\right\vert _{A}\right) \nonumber \\
&&+\nu \Delta t_{0}\frac{\alpha_{-}}{\alpha_{+}}\left( \left.
\frac{\partial W_{-}}{\partial x_{0}}\right\vert _{A}+\frac{\Delta
x_{0}}{2}\left. \frac{\partial^2W_{-}}{\partial
x_{0}^2}\right\vert _{A}\right) .\qquad \label{B-4}
\end{eqnarray}
Since for a diffusion process $x(t)$: $\Delta x_{0}\sim \sqrt
{D\Delta t_{0}}$, from Eq.~(\ref{B-4}) we obtain in the limit
$\Delta t_{0}\rightarrow 0$
\begin{equation}
\left. \frac{\partial W_{+}}{\partial x_{0}}\right\vert _{A}=0.
\label{B-5}
\end{equation}
By differentiating Eq.~(\ref{warthe}) (see Appendix A) and using
Eq.~(\ref{B-5}) we find
\begin{equation}
T_{+}^{\prime}\left( A\right)=0, \quad T_{-}^{\prime}\left(
A\right)=0, \label{B-6}
\end{equation}
that are the boundary conditions (\ref{cond-1}) used in our paper.

If we request an immediate switching of dichotomous noise $\eta
(t)$ when Brownian particle reaches the reflecting boundary $A$,
we must put $\beta (\Delta t_{0})=1$ in Eq.~(\ref{B-3}). As a
result, we have
\begin{eqnarray}
-\Delta t_{0}\frac{\partial W_{+}}{\partial t_{0}} &\simeq&
\alpha_{-}\Delta x_{0}\left( \left. \frac{\partial W_{-}}{\partial
x_{0}}\right\vert _{A}+\frac{\Delta x_{0}}{2}\left.
\frac{\partial^2W_{-}}{\partial x_{0}^2}\right\vert _{A}\right)
\nonumber \\
&+&W_{-}-W_{+}.
\label{B-7}
\end{eqnarray}
In the limit $\Delta t_{0}\rightarrow 0$ we obtain from
Eq.~(\ref{B-7})
\begin{equation}
W_{+}=W_{-}.
\label{B-8}
\end{equation}
Taking into account equality (\ref{B-8}) in Eq.~(\ref{B-7}) and
dividing both sides of this equation by $\alpha_{-}\Delta x_{0}$,
we find in the limit $\Delta t_{0}\rightarrow 0$
\begin{equation}
\left. \frac{\partial W_{-}}{\partial x_{0}}\right\vert _{A}=0.
\label{B-9}
\end{equation}
Equations (\ref{B-8}) and (\ref{B-9}) are equivalent to the
following conditions for MFPTs at the reflecting boundary
\begin{equation}
T_{+}\left( A\right)=T_{-}\left( A\right) , \quad
T_{+}^{\prime}\left( A\right)=0 \label{B-10}
\end{equation}
and
\begin{equation}
T_{+}\left( A\right)=T_{-}\left( A\right) , \quad
T_{-}^{\prime}\left( A\right)=0. \label{B-11}
\end{equation}
It would be emphasized that conditions (\ref{B-10}) and
(\ref{B-11}) were previously derived in Ref.~\cite{Bal88} by more
complex procedure. We must choose the conditions (\ref{B-10}) or,
alternatively, the conditions (\ref{B-11}) from physical
considerations.

\section{Appendix C}

Using Eqs.~(\ref{third}) we can eliminate the unknown constant
$c_6$ from Eq.~(\ref{T-}) and after rearrangements obtain
\begin{eqnarray}
T_{\pm}\left( 0\right) &=&\frac{b}{k}+\frac{\nu L^{2}}{\Gamma
^{2}} + \sum_{i=1}^{3} ac_{i}\left( \frac{e^{\mu
_{i}}}{k+\lambda _{i}D}+\frac{D\lambda_{i}}{\Gamma ^{2}}\right) \nonumber \\
-\frac{\Gamma ^{2}}{2\nu k^{2}}&-&c_{5}\frac{2\nu L}{a} \mp
c_{5}\frac{\Gamma ^{2}}{a^2}\left( 1 \pm \frac{a}{k}\right) .
\label{Expr-MFPTs}
\end{eqnarray}
The unknown constants $c_1,c_2,c_3,c_5$ involved in
Eq.~(\ref{Expr-MFPTs}) must be found from the following set of
equations (see Eqs.~(\ref{third}))
\begin{eqnarray}
&&\sum_{i=1}^{3}c_{i}e^{\mu _{i}} = \frac{a}{2\nu k},
\nonumber\\
&&\sum_{i=1}^{3}c_{i} - c_{5}\left( 1+\frac{2\nu D}{a^{2}}\cosh
{\gamma L}\right) \nonumber \\
&& = \frac{aD}{\Gamma ^{3}}\sinh {\gamma L}+ \frac{a}{2\nu
k}-\frac{aL}{\Gamma
^{2}} , \nonumber\\
&&\sum_{i=1}^{3}c_{i}\lambda _{i} - c_{5}\frac{2\nu
\Gamma}{a^{2}}\sinh {\gamma L} = \frac{a}{\Gamma^{2}}\left( \cosh
{\gamma L}-1\right) , \nonumber \\
&&\sum_{i=1}^{3} \frac{c_{i}\lambda _{i}}{k+\lambda _{i}D} -
c_{5}\frac{2\nu }{a^2}\left( \cosh {\gamma L}-1\right) \nonumber \\
&& = \frac{2\nu L}{a\Gamma ^{2}}+\frac{a}{\Gamma ^{3}} \sinh
{\gamma L} -\frac{1}{ak}. \label{c1-c5}
\end{eqnarray}
To investigate the NES phenomenon, we look for asymptotic
expressions of Eqs.~(\ref{Expr-MFPTs}) and (\ref{c1-c5}) for
$D\rightarrow 0$, with an accuracy of linear terms. To do this we
need approximate expressions of the cubic equation roots. From
Eq.~(\ref{roots}) we have
\begin{eqnarray}
&& \lambda _{1}+\lambda _{2}+\lambda _{3}=-\frac{2k}{D},  \nonumber \\
&& \lambda _{1}\lambda _{2}\lambda _{3}=\frac{2\nu k}{D^2},  \label{sum}
\end{eqnarray}
and, as a consequence, we can put the roots in the following form
\begin{equation}
\lambda \simeq \frac{A_{1}}{D}+A_{2}+A_{3}D, \label{lambda_i}
\end{equation}
with the unknown constants $A_{1},A_{2},A_{3}$. Substitution of
Eq. (\ref{lambda_i}) in Eq.~(\ref{roots}) gives
\begin{eqnarray}
&& \lambda _{1}\simeq -\frac{k+a}{D}-\frac{\nu }{k+a},\nonumber \\
&& \lambda _{2}\simeq -\frac{k-a}{D}-\frac{\nu }{k-a},\nonumber \\
&& \lambda _{3}\simeq \frac{2\nu k}{k^{2}-a^{2}}\left[
1-\frac{2\nu D\left( k^{2}+a^{2}\right) }{\left(
k^{2}-a^{2}\right) ^{2}}\right] .  \label{apro}
\end{eqnarray}
As it is seen from Eq.~(\ref{apro}), in the limit $D\rightarrow 0$
($\gamma \rightarrow +\infty $) we have: $\mu _{1}\rightarrow
-\infty $ and $\mu _{2}\rightarrow -\infty $. Therefore, the
negligibly small terms with $e^{\mu _{1}}$ and $e^{\mu _{2}}$ can
be neglected in Eq.~(\ref{Expr-MFPTs}) and in the first equation
(\ref{c1-c5}), and we find
\begin{equation}
c_{3} \simeq \frac{a}{2\nu k}\cdot e^{-\mu _{3}}. \label{c3}
\end{equation}
Using the following approximate expressions
$$
\sinh \gamma L\simeq \cosh \gamma L\simeq \frac{e^{\gamma
L}}{2}\qquad \left( \gamma \rightarrow +\infty \right) ,
$$
we find $c_5$ from Eqs.~(\ref{c1-c5})
\begin{equation}
c_5 \simeq -\frac{a^3}{2\nu \Gamma^{3}}. \label{c5}
\end{equation}
Substituting Eqs.~(\ref{c3}) and (\ref{c5}) in
Eq.~(\ref{Expr-MFPTs}), and solving the system (\ref{c1-c5}), we
obtain
\begin{eqnarray}
&& T_{\pm}\left( 0\right) \simeq\frac{b}{k}+\frac{\nu
L^{2}}{\Gamma ^{2}} + \frac{a^2}{2\nu k\left( k+\lambda
_{3}D\right)}
\nonumber \\
 &&  + \frac{a^2[k(\Gamma^2-4\nu D+2\nu L\Gamma)-
\Gamma^2(\Gamma + D\lambda_3)]}{\Gamma^4 \lambda_3(k + D\lambda_3)
(k + D\lambda_3 + \Gamma)} \nonumber \\
&&  + \frac{a^2e^{-\mu_3} [(k + D \lambda_3)(k + 3D \lambda_3) -
\Gamma^2]}{2\nu k \Gamma (k + D\lambda_3)
(k + D\lambda_3 + \Gamma)} \nonumber \\
&& + \frac{a^{2} L}{\Gamma^{3}} -\frac{\Gamma ^{2}}{2\nu k^{2}}
\pm \frac{a}{2\nu \Gamma }\left( 1 \pm \frac{a}{k}\right) .
\label{Final MFPTs}
\end{eqnarray}
After expanding the expression (\ref{Final MFPTs}) in power series
in $D$ up to linear terms we derive the main result
(\ref{T-approx})-(\ref{main}).


\begin{thebibliography}{30}

\bibitem{Rei02}
P. Reimann, Phys. Rep. \textbf{361}, (2002) 57-265.

\bibitem{Sme99}
V. N. Smelyanskiy, M. I. Dykman, and B. Golding, Phys. Rev. Lett.
\textbf{82}, (1999) 3193-3197.

\bibitem{Leh00}
J. Lehmann, P. Reimann, and P. H\"{a}nggi, Phys. Rev. Lett.
\textbf{84}, (2000) 1639-1642.

\bibitem{Mai01}
R. S. Maier and D. L. Stein, Phys. Rev. Lett. \textbf{86}, (2001)
3942-3945.

\bibitem{Hir82}
J. E. Hirsch, B. A. Huberman, D. J. Scalapino, Phys. Rev. A
\textbf{25}, (1982) 519-532.

\bibitem {Day92}
I. Dayan, M. Gitterman, G. H. Weiss, Phys. Rev. A \textbf{46},
(1992) 757-761.

\bibitem{Man96}
R. N. Mantegna and B. Spagnolo, Phys. Rev. Lett. \textbf{76},
(1996) 563-566.

\bibitem{Man98}
R. N. Mantegna and B. Spagnolo, Int. J. Bifurcation and Chaos
\textbf{8}, (1998) 783-790.

\bibitem{Man00}
R. N. Mantegna and B. Spagnolo, \emph{Stochastic Processes in
Physics, Chemistry and Biology, Lecture Notes in Physics} (Eds. J.
A. Freund  and T. Poeschel, Springer-Verlag, Berlin 2000) 327-337.

\bibitem{Agu01}
N. V. Agudov and B. Spagnolo, Phys. Rev. E \textbf{64}, (2001)
035102(R)-1 - 035102(R)-4.

\bibitem{Agu00}
N. V. Agudov and B. Spagnolo, \emph{Stochastic and Chaotic
Dynamics in the Lakes} (Eds. Broomhead D. S., et al, American
Institute of Physics, 2000) 272-277.

\bibitem{Agu03}
N. V. Agudov, A. A. Dubkov, and B. Spagnolo, Physica A
\textbf{325}, (2003) 144-151.

\bibitem{Fia03}
A. Fiasconaro, D. Valenti, and B. Spagnolo, Physica A
\textbf{325}, (2003) 136-143.

\bibitem{Mal96}
A. N. Malakhov, A. L. Pankratov, Physica C \textbf{269}, (1996)
46-54.

\bibitem{Spa04}
B. Spagnolo, A. A. Dubkov, and N. V. Agudov, Acta Physica Polonica
B \textbf{35}, (2004) 1419-1436.

\bibitem{Wac98}
R. Wackerbauer, Phys. Rev. E \textbf{58}, (1998) 3036-3044.

\bibitem{Wac99}
R. Wackerbauer, Phys. Rev. E \textbf{59}, (1999) 2872-2879.

\bibitem{Mie00}
A. Mielke, Phys. Rev. Lett. \textbf{84}, (2000) 818-821.

\bibitem{Dan99}
D. Dan, M. C. Mahato, and A. M. Jayannavar, Phys. Rev. E
\textbf{60}, (1999) 6421-6428.

\bibitem{Mah97}
M. C. Mahato and A. M. Jayannavar, Mod. Phys. Lett. B \textbf{11},
(1997) 815-820.

\bibitem{Mah98}
M. C. Mahato and  A. M. Jayannavar, Physica A \textbf{248}, (1998)
138-154.

\bibitem{Mat83}
K. Matsumoto and J. Tsuda, Phys. Rev. A \textbf{32}, (1985)
1934-1937.

\bibitem{Yos03}
M. Yoshimoto, Phys. Lett. A \textbf{312}, (2003) 59-64.

\bibitem{Xie03}
Xie Chong-Wei and Mei Dong-Cheng, Chin. Phys. Lett. \textbf{20},
(2003) 813-816.

\bibitem{Apo97}
F. Apostolico, L. Gammaitoni, F. Marchesoni, and S. Santucci,
Phys. Rev. E \textbf{55}, (1997) 36-39.

\bibitem{Doe92}
C. R. Doering and J. C. Gadoua, Phys. Rev. Lett. \textbf{69},
(1992) 2318-2321.

\bibitem{Mar96}
M. Marchi, F. Marchesoni, L. Gammaitoni, E. Menichella-Saetta, and
S. Santucci, Phys. Rev. E \textbf{54}, (1996) 3479-3487.

\bibitem{ManSpa00}
R. N. Mantegna and B. Spagnolo, Phys. Rev. Lett. \textbf{84},
(2000) 3025-3028.

\bibitem{Han85}
P. H\"anggi and  P. Talkner, Phys. Rev. A \textbf{32}, (1985)
R1934-R1937.

\bibitem{Mas86}
J. Masoliver, K. Lindenberg, and B. J. West, Phys. Rev. A
\textbf{33}, (1986) 2177-2180.

\bibitem{Rod86}
M. A. Rodrigues and L. Pesquera, Phys. Rev. A \textbf{34}, (1986)
4532-4534.

\bibitem{Doe87}
C. R. Doering, Phys. Rev. A \textbf{35}, (1987) 3166-3167.

\bibitem{Bal88}
V. Balakrishnan, C. Van den Broeck, and P. H\"{a}nggi, Phys. Rev.
A \textbf{38}, (1988) 4213-4222.

\bibitem{Zur93}
U. Z\"{u}rcher and C. R. Doering, Phys. Rev. E \textbf{47}, (1993)
3862-3869.

\bibitem{Luc92}
J. \L uczka, M. Niemiec, and E. Piotrowski, Phys. Lett. A
\textbf{167}, (1992) 475-478.

\bibitem{Dub04}
A. A. Dubkov, N. V. Agudov, and B. Spagnolo, Phys. Rev. E
\textbf{69}, (2004) 061103-1~-~061103-7.

\bibitem{Fur63}
K. Furutsu, J. Res. Nat. Bur. Stand. \textbf{67D}, (1963) 303-323.

\bibitem{Nov65}
E. A. Novikov, Sov. Phys. JETP \textbf{20}, (1965) 1290-1294.

\bibitem{Sha78}
V. E. Shapiro and V. M. Loginov, Physica A \textbf{91}, (1978)
563-574.

\end{thebibliography}
\end{document}